\documentclass[]{spie}

\pdfoutput=1 

\usepackage[]{graphicx,color}
\usepackage[]{amsmath}
\usepackage[]{amssymb}
\usepackage[]{deluxetable}
\usepackage{float}
\usepackage{placeins}

\def\gtrsim{\mathrel{\hbox{\rlap{\hbox{\lower4pt\hbox{$\sim$}}}\hbox{$>$}}}}

\title{The Planetary Systems Imager: 2-5 Micron Channel}

\author{
Andrew J. Skemer\authorinfo{\supit{*} Contact: askemer@ucsc.edu}\supit{a},
Deno Stelter\supit{a},
Dimitri Mawet\supit{b},
Michael Fitzgerald\supit{c},
Benjamin Mazin\supit{d},
Olivier Guyon\supit{e},
Christian Marois\supit{f},
Zackery Briesemeister\supit{a},
Timothy Brandt\supit{e},\\
Jeffrey Chilcote\supit{g},
Jacques-Robert Delorme\supit{b},
Nemanja Jovanovic\supit{b},
Jessica Lu\supit{h},
Maxwell Millar-Blanchaer\supit{i},
James Wallace\supit{i},
Gautam Vasisht\supit{i},
Lewis C. Roberts Jr. \supit{i}
and
Ji Wang\supit{b}
\skiplinehalf
\supit{a} University of California, Santa Cruz, USA;
\skiplinehalf
\supit{b} California Institute of Technology, USA;
\skiplinehalf
\supit{c} University of California, Los Angeles, USA;
\skiplinehalf
\supit{d} University of California, Santa Barbara, USA;
\skiplinehalf
\supit{e} National Astronomical Observatory of Japan, Japan;
\skiplinehalf
\supit{f} NRC-Dominion Astrophysical Observatory, Canada;
\skiplinehalf
\supit{g} Stanford University, USA;
\skiplinehalf
\supit{h} University of California, Berkeley, USA;
\skiplinehalf
\supit{i} Jet Propulsion Lab, California Institute of Technology, USA;
}

\begin{document} 
\maketitle

\begin{abstract}
We summarize the red channel (2-5 micron) of the Planetary Systems Imager (PSI), a proposed second-generation instrument for the TMT.  Cold exoplanets emit the majority of their light in the thermal infrared, which means these exoplanets can be detected at a more modest contrast than at other wavelengths.  PSI-Red will be able to detect and characterize a wide variety of exoplanets, including radial-velocity planets on wide orbits, accreting protoplanets in nearby star-forming regions, and reflected-light planets around the nearest stars.  PSI-Red will feature an imager, a low-resolution lenslet integral field spectrograph, a medium-resolution lenslet+slicer integral field spectrograph, and a fiber-fed high-resolution spectrograph.
\end{abstract}

\section{INTRODUCTION\label{sec:intro}}
Extremely Large Telescopes (ELTs) will have the angular resolution to image a variety of exoplanets that are closer to their host stars than the exoplanets that can be imaged with current facilities.  The Planetary System Imager (PSI) is proposed instrument that is designed to image exoplanets at a variety of wavelengths and spectral resolutions\cite{Fitzgerald_SPIE2018}.  Figure 1 shows	a possible architecture for PSI.  Light from the telescope is corrected by an adaptive optics (AO) system, featuring a large-format woofer deformable mirror\cite{Guyon_SPIE2018}.  The AO system comprises a single relay with gold-coated surfaces to limit thermal emissivity.  Light from the adaptive optics system is directed to the cryogenic 2-5 micron channel (PSI-Red), and a tilted dichroic entrance window transmits the $>$2 micron light while reflecting $<$2 micron light.  The shorter wavelength light is used in a downstream wavefront sensor that controls the woofer AO system.  Additionally some of the shorter wavelength light is further corrected by a tweeter AO system\cite{Guyon_SPIE2018} and directed to a short wavelength science camera (PSI-Blue\cite{Mawet_SPIE2018}).  PSI-Red will directly detect and characterize thermal emission from exoplanets while PSI-Blue will detect and characterize reflected light from exoplanets.

\begin{figure}[htbp]
\begin{center}
  \hbox{
    \hspace{0.1in}
      \includegraphics[angle=0,width=1.0\linewidth]{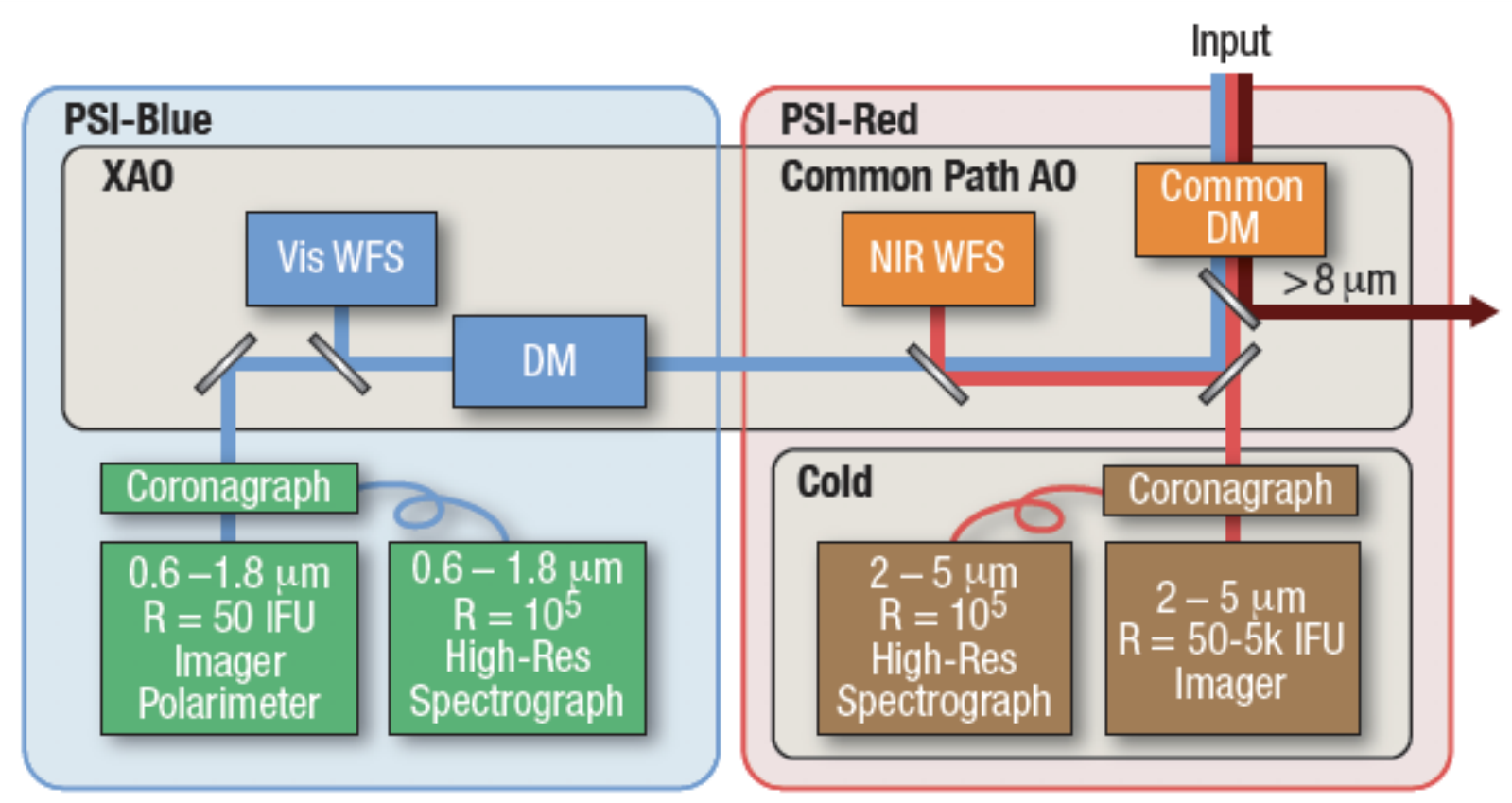}}
\end{center}
      \vspace{0.0in}
      \caption{--- Schematic of the Planetary Systems Imager\cite{Fitzgerald_SPIE2018}.  Light from a high-order woofer AO system\cite{Guyon_SPIE2018} passes an insertable pickoff dichroic to a 10 $\mu$m camera\cite{Marois_SPIE2018}.  Light enters the PSI-Red science camera via a dichroic entrance window that transmits 2-5 $\mu$m light and reflects shorter wavelength light to a near-infrared wavefront sensor and PSI-blue\cite{Mawet_SPIE2018}.
	  } \label{fig:PSI schematic}
\end{figure}

PSI-Red comprises an imager, a low-resolution integral-field spectrograph, a medium-resolution integral-field spectrograph, and a fiber-injection unit for high-resolution spectroscopy.  Each of these channels, as well as a shared set of coronagraphic foreoptics, is housed in a cryogenic dewar to minimize the thermal infrared background.  PSI-Red is a diffraction-limited instrument, which has an optical design and instrument volume that are independent of telescope aperture.  As a result, we are planning a precursor instrument for Keck, SCALES (Santa Cruz Array of Lenslets for Exoplanet Spectroscopy), which is identical to the current PSI-Red design.  SCALES could, in principle, be used on Keck while the TMT is under construction, and then integrated into the rest of PSI as a finished product. 

\section{IMAGING EXOPLANETS WITH PSI-RED\label{sec:exoplanets}}
There are three major categories of open questions about exoplanets:

\begin{enumerate}
	\item{\textbf{Planet Demographics}} What is the frequency of planets as a function of separation and mass?  What is the complete set of possible outcomes from the planet formation process?
	\item{\textbf{Planetary System Architectures}} What correlations exist between the locations and properties of different planets in a planetary system?  
	\item{\textbf{Planet Characterization}} What are the properties of individual planets, and how do these properties vary among the full set of new and diverse worlds?
\end{enumerate}

PSI's scientific focus will primarily be items 2 and 3.  Various radial velocity surveys and space telescopes like \textit{Kepler}, \textit{TESS} and \textit{WFIRST} are putting strong constraints on the frequency of planets as a function of separation and mass\cite{2013ApJS..204...24B,2015JATIS...1a4003R,2014arXiv1409.2759Y}.  However, to study exoplanets beyond their locations, masses and radii requires photometry and spectroscopy of the planets themselves.  While transit and eclipse spectroscopy is possible for relatively hot inner planets, direct imaging and spectroscopy is necessary for studying cold planets, planets on long-period orbits, planets with thin atmospheres, and planetary surfaces.

PSI-Red will initially focus on studying the properties of gas-giant exoplanets on wide period orbits.  As the AO performance of PSI is refined, it will become possible to image lower-mass planets at smaller separations, including warm super-Earths and mini-Neptunes.\cite{2015IJAsB..14..279Q}

PSI-Red combines two of the most successful methods for imaging exoplanets, thermal infrared imaging and integral field spectroscopy, which will allow the detection of colder and lower-mass exoplanets than have been imaged before.  In terms of improving our ability to see faint exoplanets, the main benefits of combining thermal infrared imaging with integral field spectroscopy are as follows:

\begin{enumerate}
\item Planets can be imaged with more moderate planet-star contrasts in the thermal infrared than in the near-infrared (see Figure 2)\cite{2014ApJ...792...17S}
\item With broad-bandpass integral field spectroscopy, speckles can be distinguished from astrophysical sources because they move radially as a function of wavelength, while planets stay fixed.\cite{2006SPIE.6272E..0LM}
\item Speckles can be distinguished from planets via their spectroscopic signatures because planets have large molecular absorption features that do not exist in stellar spectra (see Figures 2 and 3)\cite{2007ApJS..173..143B}
\item The optimal bandpass for detecting exoplanets varies dramatically as a function of exoplanet temperature, and can be adjusted in post-processing with integral field spectroscopy (see Figure 3). This is much more important in the thermal infrared than in the near-infrared.\cite{2016ApJ...817..166S}
\end{enumerate}

\begin{figure}[htbp]
\begin{center}
  \hbox{
    \hspace{1.4in}
      \includegraphics[angle=0,width=0.7\linewidth]{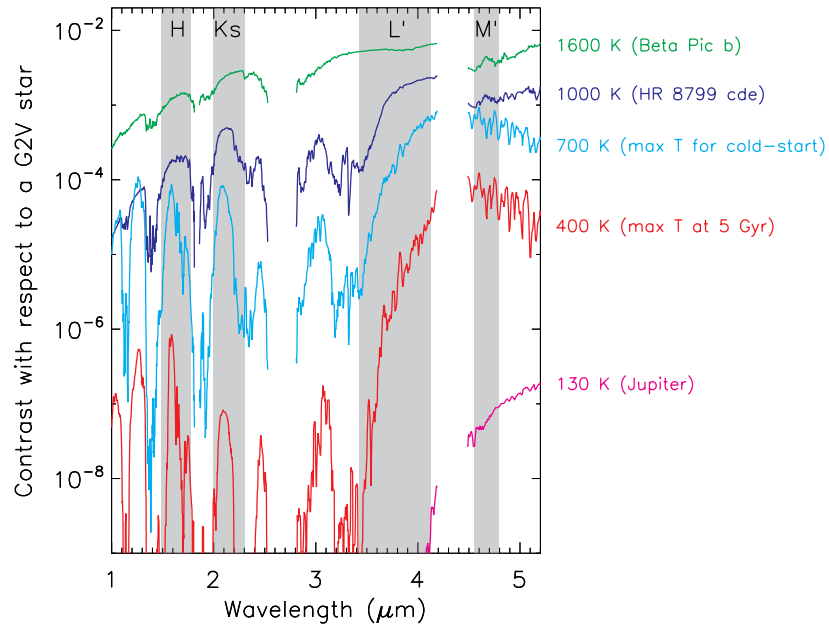}}
\end{center}
      \vspace{-0.1in}
      \caption{--- Self-luminous exoplanets of all temperatures can be imaged at lower contrasts in the thermal infrared (3-5 $\mu$m) than in the near infrared (1-2 $\mu$m). This becomes more pronounced at lower temperatures.\cite{2014ApJ...792...17S}
      } \label{fig:L contrast}
\end{figure}

\begin{figure}[htbp]
\begin{center}
  \hbox{
    \hspace{0.2in}
      \includegraphics[angle=0,width=0.9\linewidth]{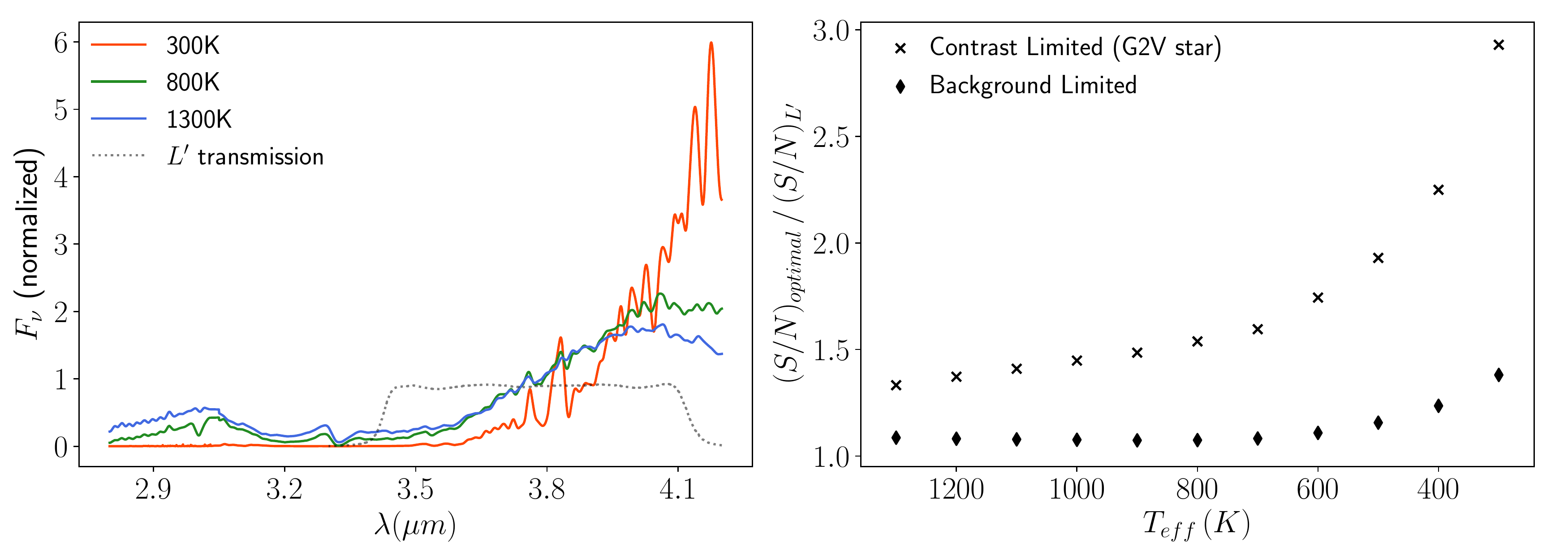}}
\end{center}
      \vspace{-0.3in}
      \caption{--- Left: Exoplanets SEDs vary dramatically with wavelength so that a single standard filter (such as L’) is not optimal for a wide range of planets.  Right: Using an IFS, the optimal bandpass can be selected.  In the sky-background-limited regime, a 300 K can be detected with 40\% more S/N by using a weighted average filter instead of a standard L’ filter.  In the contrast-limited regime, a 300 K planet can be detected with 200\% more S/N by using a weighted average filter instead of a standard L’ filter.  This is independent of any speckle suppression, where in IFS also helps.  The combination of an optimal thermal infrared filter and speckle suppression from an IFS creates a ~4-5 magnitude boost in sensitivity compared to an H-band IFS and a ~2.2 magnitude boost in sensitivity compared to an L-band imager.
	  } \label{fig:cold planet boost}
\end{figure}

\section{DESIGN\label{sec:design}}
PSI-Red is a diffraction-limited 2-5$\mu$m instrument, capable of imaging and spectroscopy over a range of spectral resolutions.  The most demanding technical specifications for PSI-Red come from its driving science case, which is discovering and characterizing exoplanets with low-resolution integral field spectroscopy.  PSI-Red is designed to maximize image quality (contrast) and sensitivity with the low-resolution IFS.  Various additional modules are all designed to be passive with respect to the low-resolution IFS.  We describe each module below.  

\textbf{Dewar}---In its current design, the PSI-Red dewar is 1580 x 1180 x 900 mm and weighs 400 kg.  Its optical bench is cooled to 70 K and its detectors are cooled to 55 K by two closed-cycle coolers.  The top and bottom of the PSI-Red dewar are both removable, revealing an optics bench that is populated on both sides.  All hermetic feedthroughs go through the sides of the dewar so the lids can be removed without having to unattach cables.  

\textbf{Low Resolution IFS, including foreoptics}---Light from PSI-AO enters the PSI-Red dewar through a CaFl entrance window and goes through the foreoptics on the bottom of the optics plate.  The AO focus is reimaged by an OAP relay, which creates a pupil plane containing a fixed cold-stop, and a focal plane containing a wheel with coronagraphs optimized for K-, L-, and M-bands.  A second OAP relay magnifies the image by 22.8 to create a suitable plate-scale for the IFS lenslet array.  This relay creates a second pupil plane where a wheel holds various Lyot stops matched to the coronagraphs.  After reflecting off of two flats, the beam emerges on the top of the PSI-Red optical bench and creates an image on the lenslet array.  Each lenslet is a 341$\mu$m square, which samples the field with a 20 mas plate-scale.  The lenslet array is photolithographically etched on silicon (based on the design for ALES\cite{2015SPIE.9605E..1DS,Skemer_2018SPIE_ALES}) and creates an f/8 beam for each lenslet (this f/8 beam size was chosen to compromise between aberrations in the spectrograph and diffraction from the lenslets).  The lenslet beams go through a three-mirror-anastigmat collimator, which produces a pupil plane where a wheel is populated with reflective Sapphire-ZnSe prism pairs that each correspond to a fixed bandpass.  The dispersed collimated light is imaged onto an H4RG detector with a three-mirror-anastigmat camera.  At the detector, PSI-Red creates a 180x180 array of spectra, each of which is 72-pixels long and separated from neighboring spectra by 7.2 pixels.  Other than the entrance window, the lenslet array, and the prisms, PSI-Red has an all-reflective design that can be aligned at room temperature.

\begin{figure}[htbp]
\begin{center}
  \hbox{
    \hspace{0.2in}
      \includegraphics[angle=0,width=0.9\linewidth]{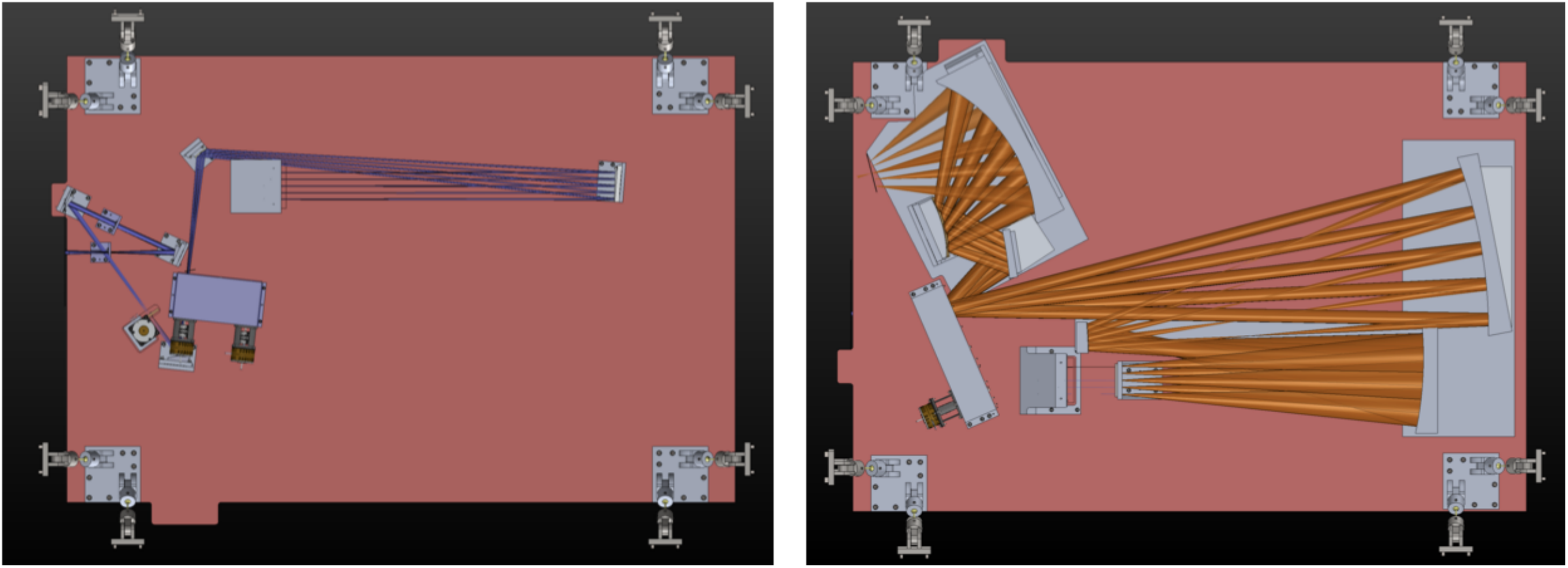}}
\end{center}
      \vspace{-0.3in}
      \caption{--- Schematic of the foreoptics (left) and integral field spectrograph (right) of PSI-Red / SCALES.  Light enters the dewar from the left, and goes through two relays, which contain a cold-stop, coronagraph, and Lyot stop, as the beam is magnified to the lenslet array plate scale.  The silicon lenslet array  images 1-to1 onto an H4RG detector, via a three-mirror-anastigmat collimator and camera.  The light is dispersed via selectable reflective prisms.
	  } \label{fig:low-res}
\end{figure}

\textbf{Medium Resolution IFS}---After the lenslet array, an insertable fold mirror picks off light from the central 23x23 spaxels (0.46x0.46”) of the low-resolution IFS’s field-of-view.  The lenslet spots are imaged onto a 23-element image slicer, which sends the beams to an 8-element pupil re-imager, and another insertable fold that passes the 529 spots back into the spectrograph as a single pseudo-slit with a length of 3800 pixels on an H4RG.  Gratings in the collimated pupil wheel disperse perpendicular to the slit up to a length of ~4000 pixels (baselined as one each for K-, L-, and M-bands).  Normally, image slicers create aberrations that are not amenable to the stringent constraints of exoplanet imaging.  However, because this slicer is inserted after the lenslet array, the field has already been sampled and as long as the spots remain separated, the slicer has no effect on image quality, similar to the Keck Da Vinci concept\cite{2010SPIE.7735E..7SA}.  In the case where a star is placed on the PSI-Red coronagraph and the planet is moving in the field while PSI-AO tracks the pupil, the flat mirror that sends light to the top of the optics plate will be steered with a low-hysteresis, cryogenic piezo-electric device. 

\textbf{Imager}---The Lyot stop wheel in the fore-optics will contain a mirror and beamsplitters that divert light to an imaging camera (reflective Lyot masks will also use the imager for coronagraph alignment monitoring).  An additional OAP relay reimages a 7x7” field-of-view (with 3.5 millisarcsecond sampling) onto an H2RG detector.  Various bandpass filters are inserted in a multi-plane wheel at a location near the PSI-Red cold-stop.  PSI-Red is all reflective through the imaging channel, other than its dichroic entrance window.

\textbf{Focal-Plane Wavefront Sensing}---The second fold that brings the light from the bottom of the optics plate to the top can be replaced by 4 beam-splitters in a wheel that reflect light to the lenslet array and pass light to a zero-read-noise Saphira infrared detector that is used as a focal-plane wavefront sensor for speckle nulling.  This design minimizes non-common-path aberrations between the science field (sampled at the lenslet array) and the wavefront sensor by having just a single flat beamsplitter defining the differing optical paths.  Additionally, the beams at this point are f/340 which mitigates the slight non-common-path aberrations from the beam-splitter.

\textbf{Polarimetry}---Imaging polarimetry is easily accomplished by including a Wollaston prism in the prism wheel of the integral field spectrograph.  A rotating half-waveplate will be installed in front of the PSI-AO system

\textbf{Fiber-Injection Unit}---A fiber injection unit modeled after the Keck Planet Imager and Characterizer (KPIC\cite{2017SPIE10400E..29M,Mawet_SPIE2018_KPIC}) will allow high-resolution spectroscopy of exoplanets after PSI-Red's cryogenic fore-optics and coronagraphs.  Design work is at an early stage, but there is space allocated. on the bottom of the PSI-Red optics bench.

A more detailed design of PSI-Red is presented in Stelter et al. (these proceedings)\cite{Stelter_SPIE2018}.

\begin{figure}[htbp]
\begin{center}
  \hbox{
    \hspace{0.2in}
      \includegraphics[angle=0,width=0.9\linewidth]{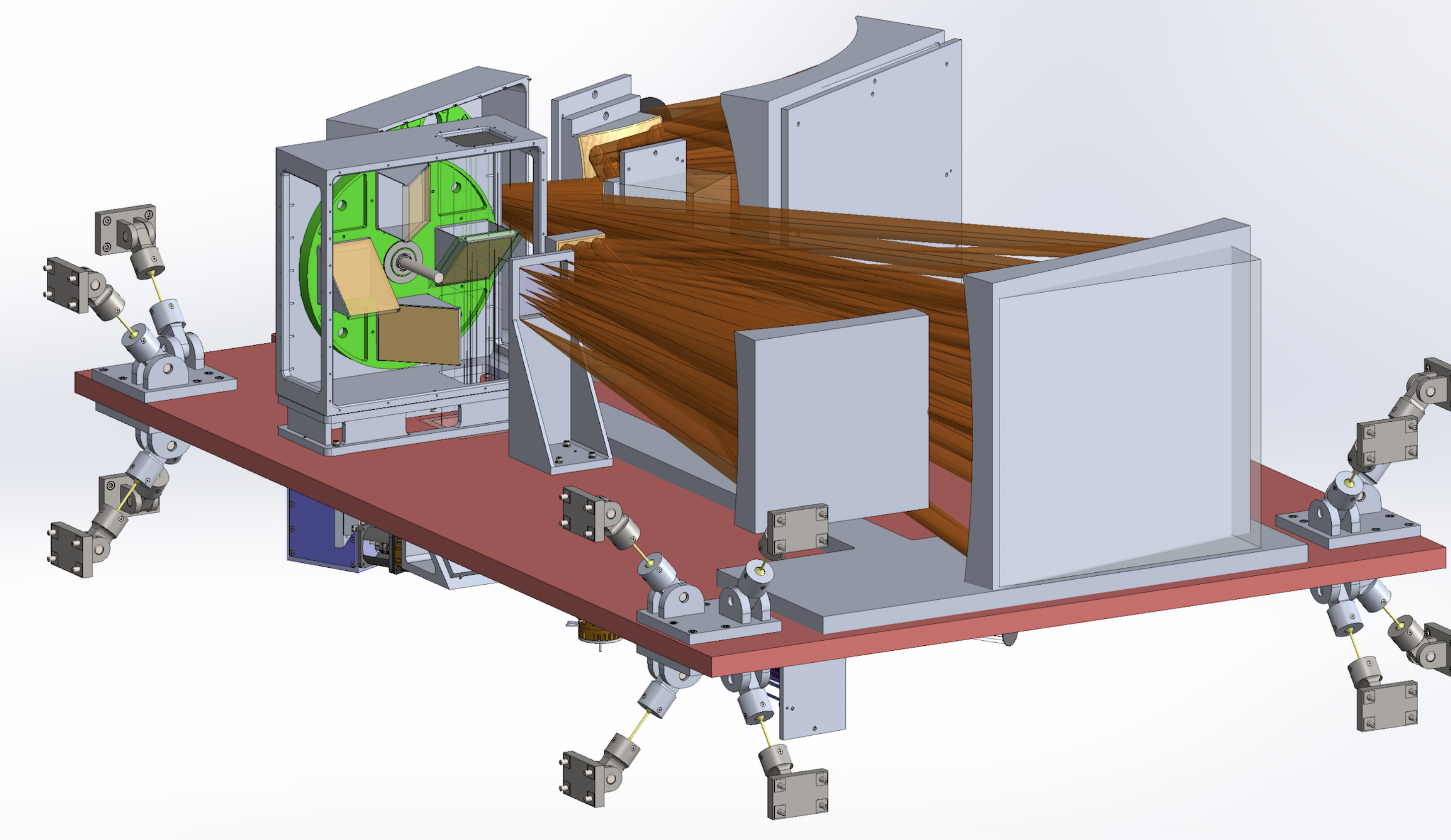}}
\end{center}
      \vspace{-0.3in}
      \caption{--- An isometric view of the PSI-Red/SCALES optical bench.
	  } \label{fig:iso}
\end{figure}

\section{SPECIFICATIONS\label{sec:specifications}}
Table 1 lists PSI-Red's sensitivities, spectral resolutions, sampling, and fields-of-view for its various modes. 

\begin{figure}[htbp]
\begin{center}
  \hbox{
    \hspace{0.1in}
      \includegraphics[angle=0,width=0.9\linewidth]{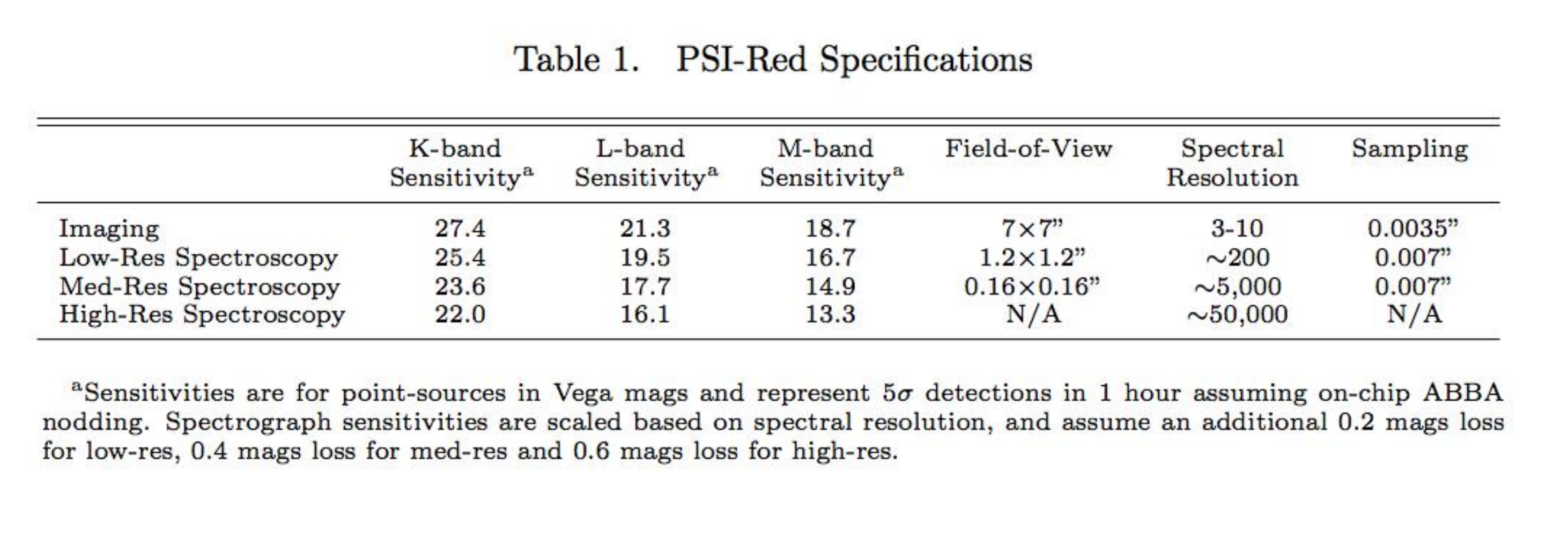}}
\end{center}
      \vspace{-0.3in}
\end{figure}


\section*{ACKNOWLEDGMENTS}
The authors wish to acknowledge the Center for Adaptive Optics Fall Retreat and the TMT Science Forum, where much of this work originated.  This paper is based on work funded by NSF Grants 1608834 and 1614320.

\bibliography{spie}
\bibliographystyle{spiebib}

\end{document}